\documentclass[sigconf, nonacm]{acmart}

\usepackage{graphicx}
\usepackage{url}
\usepackage{xcolor}
\usepackage{xspace}
\usepackage{makecell}
\usepackage{amsthm}
\usepackage{amsmath}
\usepackage{textcomp}
\usepackage{booktabs}
\newcommand{\omni}{\textsc{Sema-SQL}\xspace}
\usepackage{paralist}
\usepackage[vlined,commentsnumbered,ruled,linesnumbered]{algorithm2e}
\usepackage{enumitem}
\usepackage{graphicx}
\usepackage{subcaption}  % Add this in your preamble
\usepackage{listings}
\usepackage{xcolor}  % for colors
 \usepackage{multirow}
\usepackage{fancyvrb}

\newtheorem{definition}{Definition} [section]

\newtheorem{problem}{Problem} [section]

\usepackage{xcolor}
\usepackage{pifont}
\usepackage{graphicx}  % 用于 \rotatebox
\usepackage{xcolor}
\definecolor{darkred}{RGB}{139,0,0}

\newcommand{\correct}{{\color{green!60!black}\ding{52}}}
\newcommand{\wrong}{{\color{red!70!black}\ding{55}}}
\newcommand{\pcorrect}{{\color{cyan}\ding{52}\rotatebox[origin=c]{-9.2}{\kern-0.7em\ding{55}}}}

\usepackage{multirow}
\newcommand{\eat}[1]{{}\xspace}

\usepackage[table]{xcolor}  % includes support for table row coloring
\definecolor{bestaccuracy}{RGB}{198, 239, 206}    % Light green for best accuracy
\definecolor{lowestcost}{RGB}{189, 215, 238}      % Light blue for lowest cost
\definecolor{darkgreen}{RGB}{0, 128, 0}
\definecolor{darkblue}{RGB}{0, 70, 140}
\usepackage{xcolor}
\definecolor{omnirow}{RGB}{255, 242, 204}  % Light yellow
\definecolor{mygrey}{RGB}{230,230,240}

\begin{document}

\title{\omni: Extending Relational Queries with Large Language Models}

\author{
Yin Lin,
Tianjing Zeng,
Zhongjun Ding,
Rong Zhu,
Bolin Ding$^{*}$,
H. V. Jagadish$^{\S}$,
Jingren Zhou
}

\affiliation{%
Alibaba Group\\
$^\S$University of Michigan\\
\vspace{2mm}
\texttt{\small \{yin.lin, zengtianjing.ztj, dingzhongjun.dzj, red.zr, bolin.ding, jingren.zhou\}@alibaba-inc.com}\\
\texttt{\small jag@umich.edu}
}

\begin{abstract}
Relational databases excel at structured data analysis, but real-world queries increasingly require capabilities beyond standard SQL, such as semantically matching entities across inconsistent names, extracting information not explicitly stored in schemas, and analyzing unstructured text. While text-to-SQL systems enable natural language querying, they remain limited to relational operations and cannot leverage the semantic reasoning capabilities of modern large language models (LLMs). Conversely, recent semantic operator systems extend relational algebra with LLM-powered operations (e.g., semantic joins, mappings, aggregations), but require users to manually construct complex query pipelines.

To address this gap, we present \omni, a system that automatically answers natural language questions by generating efficient queries that combine relational operations with LLM semantic reasoning. We formalize Hybrid Relational Algebra (HRA), a declarative abstraction unifying traditional relational operators with LLM user-defined functions (UDFs). The system automates three critical aspects: (1) query generation via in-context learning that produces HRA queries with precise natural language specifications for LLM UDFs, (2) query optimization via cost-based transformations and UDF rewriting, and (3) efficient execution algorithms that reduce LLM invocations by an average of $93\%$ in semantic joins through intelligent batching.
Extensive experiments with known benchmarks, and extensions thereof, demonstrate the significant query capability improvements possible with our design.

\end{abstract}

\maketitle

\section{Introduction}

Relational database systems have long been the dominant paradigm for storing and managing structured data. 
Business analysts, data scientists, and decision-makers often need complex database queries, but many lack the required technical expertise.
Recently, systems that provide natural language interfaces to databases have become a prominent research focus, with recent advances in large language models (LLMs) showing particular promise for SQL generation \cite{DBLP:journals/corr/abs-2411-08599, DBLP:conf/nips/PourrezaR23, DBLP:conf/coling/WangR0LBCYZYSL25, DBLP:conf/iclr/PourrezaL0CTKGS25, DBLP:journals/corr/abs-2405-16755}.  

\begin{figure} [t]
    \centering
 \includegraphics[width=0.95\linewidth]{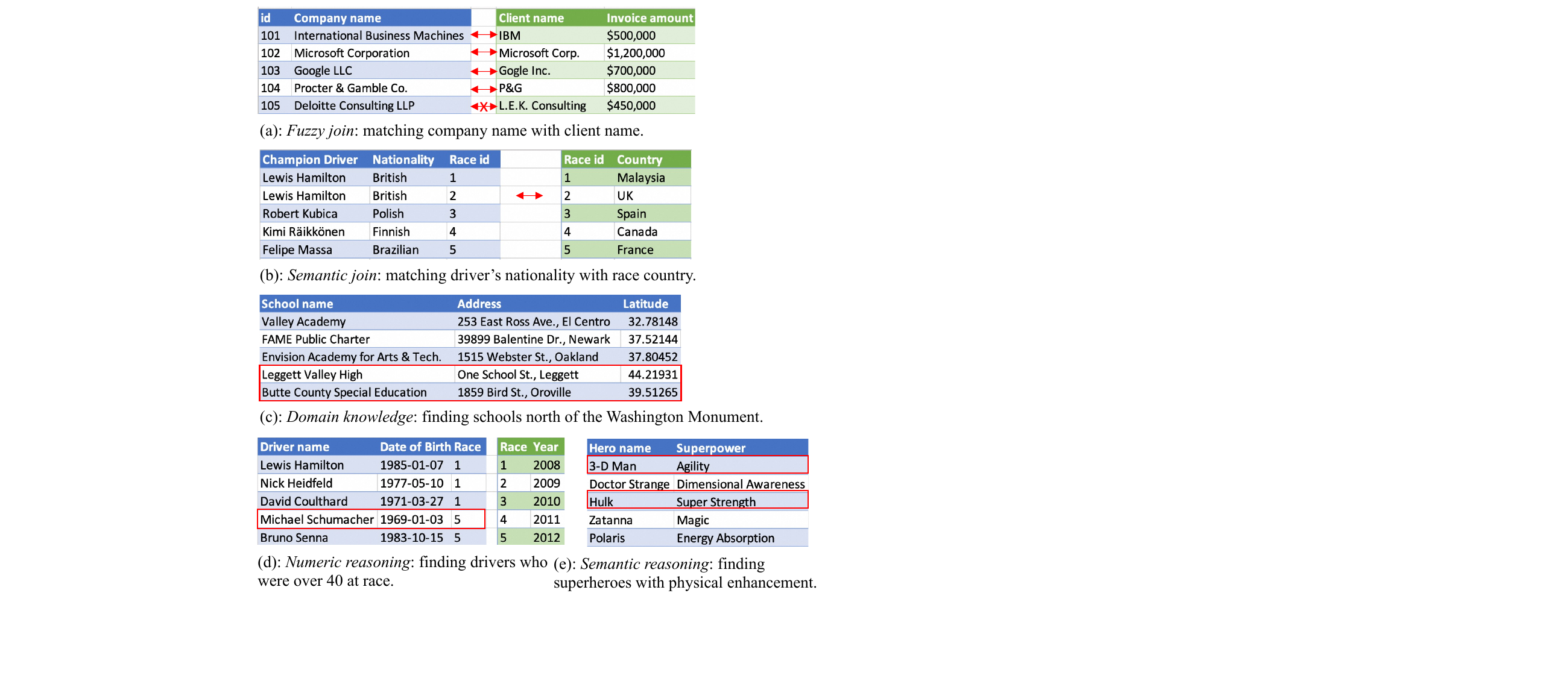}
    \vspace{-3mm}
    \caption{\textnormal{Motivating examples: extending relational querying with LLM capabilities.}}
    \vspace{-4mm}
    \label{fig:motivating}
\end{figure}

Can these approaches truly handle the complexity of real-world queries that users demand? Widely recognized text-to-SQL benchmarks such as BIRD \cite{DBLP:conf/nips/LiHQYLLWQGHZ0LC23} and Spider \cite{DBLP:conf/emnlp/YuZYYWLMLYRZR18} are constrained by the expressive limitations of relational algebra and assume that all information needed to answer user questions is contained within the original database schema. However, many real-world user queries go beyond these capabilities. Consider the following examples abridged from real-world user questions:

Figure \ref{fig:motivating}(a) shows an IT supplier querying sales revenue by customer, requiring non-equi-joins of two tables. Standard SQL using string equality matching will fail, and traditional similarity-based extensions \cite{DBLP:journals/csur/ChristophidesEP21, DBLP:journals/dke/KopckeR10} often suffer from parameter sensitivity, poor scalability, and limitations to syntactic similarity \cite{DBLP:journals/pvldb/HeGC15}.

Figure \ref{fig:motivating}(b) demonstrates a sports journalist analyzing the gap between NBA players' draft years and their first NBA MVP award. However, the draft year information is missing from the original database. SQL queries traditionally operate under the closed-world assumption, being unable to handle questions that require information not explicitly captured in the database schema \cite{corr:abs-2408-00884}.

The final example, shown in Figure~\ref{fig:motivating}(c), illustrates a restaurant owner seeking to summarize customers' favorite dishes from free-text reviews. This task requires semantic analysis of unstructured text to extract information and sentiment, extending beyond standard SQL capabilities that can only operate on structured numerical or categorical data.

\begin{figure*}[t]
    \centering
    \includegraphics[width=0.99\linewidth]{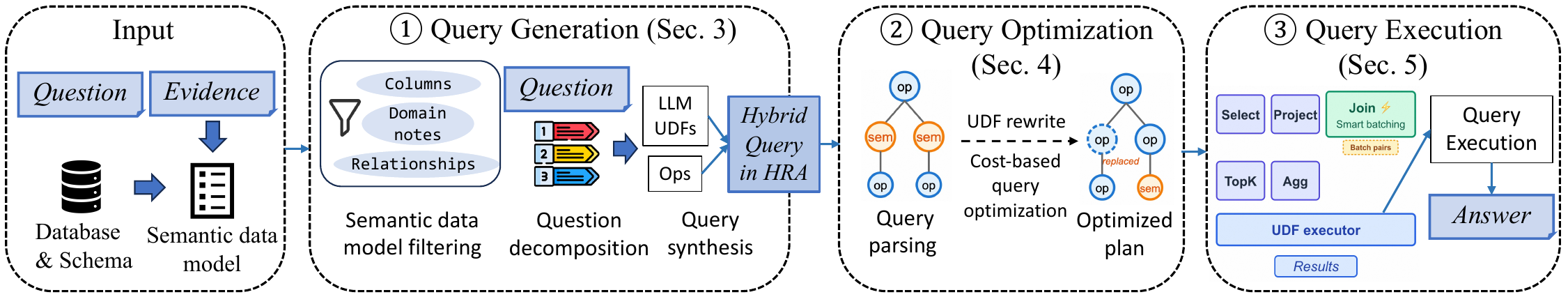}
    \vspace{-2mm}
    \caption{\textnormal{Overview of the \omni system, which operates in three phases: (1) \textit{Query Generation} translates natural language questions into HRA queries; (2) \textit{Query Optimization} optimizes query plans via a cost-based algorithm and UDF rewriting; (3) \textit{Query Execution} executes optimized plans to produce final answers.}}
    \label{fig:omni}
    \vspace{-2mm}
\end{figure*}

To support these analytical requirements, a novel class of data processing systems \cite{lotus, liu2025palimpzest, bigquery, databrick, snowflake, DBLP:journals/pacmmod/JoT24} has recently emerged that extends relational algebra with LLM capabilities through \textit{semantic operators}, including semantic filters, joins, mappings, rankings, classifications, and summarizations. These operators enable the queries shown in Figure~\ref{fig:motivating}: (a) a semantic join combines tables by matching entities in the join columns; (b) a semantic mapping extracts missing information from parametric or external knowledge; and (c) a semantic summarization performs sentiment analysis on unstructured text to infer preferences.

However, a critical challenge remains: \textit{how can we automatically translate natural language questions into efficient queries that combine both relational and semantic operations?} Existing approaches fall short in complementary ways. Hybrid question answering systems \cite{DBLP:conf/emnlp/ChenZCXWW20, stark} utilize LLMs directly to answer questions over textual and relational data, without producing queries—making answers hard to verify and results difficult to reproduce. Conversely, systems with semantic operators \cite{lotus, DBLP:conf/acl/GlennDWR24, liu2025palimpzest, DBLP:journals/pacmmod/JoT24, bigquery, databrick, redshift} require users to manually implement semantic operators and orchestrate both query construction and execution optimization—demanding expertise in both database systems and LLM-based operators.

\textbf{Our Approach.} 
We present \omni \footnote{Please check the open-sourced code: \url{https://github.com/semasql/SEMA-SQL}.}, a system that answers natural language questions over relational data by combining relational operations with LLM semantic reasoning. 
As shown in Figure \ref{fig:omni}, given a natural language question, evidence (i.e., external knowledge), and a relational database, \omni transforms the question into a declarative query in Hybrid Relational Algebra (HRA), a database-agnostic formalism that incorporates LLM-powered user-defined functions (UDFs). This query is then compiled into an optimized execution plan via cost-based transformations and UDF rewrites, then executed using specialized algorithms for semantic operators to generate the answer.

Building an end-to-end system requires addressing three key challenges. \textit{First}, LLMs are unfamiliar with semantic operators, making it challenging to synthesize queries that correctly determine when semantic operators are needed, which data they should operate on, and how to generate appropriate natural language prompts within LLM UDFs. 
To address this, \omni develops a structured prompting framework that: (1) provides a compact semantic representation of databases; (2) decomposes the query task into reasoning steps that iteratively apply \omni's relational and semantic operators; and (3) guides accurate LLM UDF generation through curated instructions and few-shot examples.

\textit{Second}, generated queries may exhibit suboptimal execution efficiency. Existing optimization approaches suffer from distinct limitations: rule-based rewriting~\cite{bigquery, DBLP:conf/acl/GlennDWR24, naacl:LiuXTSYL24, DBLP:journals/corr/abs-2508-05002} applies predetermined transformation rules (e.g., always push down predicates, always defer LLM operations) without considering actual costs; LLM-based rewriting~\cite{DBLP:journals/pvldb/ShankarCSPW25, DBLP:journals/corr/abs-2409-00847, zhu2025relationalsemanticawaremultimodalanalytics} generates plans without access to cost models or formal correctness guarantees (see Section~\ref{sec:related}). \omni extends relational query optimizers to account for LLM invocation costs—often orders 
of magnitude higher than relational operations. We develop a dynamic programming algorithm that optimally places LLM UDFs within query plans to minimize execution runtime while ensuring plan equivalence through symbolic execution~\cite{DBLP:conf/icfem/VeanesGHT09}. We further introduce UDF rewriting to replace LLM UDFs with equivalent SQL expressions where possible.

\textit{Third}, efficient execution of semantic operators remains challenging. Existing systems like LOTUS \cite{lotus} and Palimpzest \cite{liu2025palimpzest} implement semantic joins using a nested-loop approach, invoking the LLM for each row pair, which becomes prohibitively expensive. \omni\ introduces a smart-batching algorithm that dynamically groups rows from both tables into batches, adapting batch sizes to context length and task complexity. This enables the LLM to identify all matching row pairs in one invocation per batch pair, achieving significant efficiency gains while preserving accuracy.

\textbf{Contributions.}
We summarize our contributions as follows:

\begin{asparaitem}
    \item  We introduce Hybrid Relational Algebra (HRA), a formal algebraic framework that unifies relational operators with LLM-based semantic operations, providing a declarative target language for automatic query generation from natural language questions. 
    \item We develop a systematic query generation approach for HRA addressing semantic schema representation, question decomposition, and precise UDF synthesis, achieving 93.3\% query generation accuracy across three correctness criteria (syntactic validity, semantic accuracy, executability). On benchmarks requiring semantic reasoning beyond relational operations, our approach matches state-of-the-art systems that rely on manually constructed pipelines.
    
    \item We propose a cost-based optimization framework that integrates LLM invocation costs with traditional relational metrics, featuring an algorithm with equivalence guarantees for optimal UDF placement and automatic rewriting of LLM UDFs into equivalent SQL operations. Our optimizations reduce execution runtime by 28\% and token consumption by 21\%.
            
    \item We design specialized execution algorithms that integrate LLM UDF execution into database engines, including a batching optimization for semantic joins that reduces LLM invocations by $5$---$300\times$ without accuracy loss.

\end{asparaitem}

\section{Hybrid Relational Algebra}\label{sec:pre}

We introduce Hybrid Relational Algebra (HRA), which extends relational algebra with LLM-based semantic operations. HRA provides a declarative algebraic representation for queries that combine structured relational data processing with semantic reasoning. 

\begin{figure}[t]
\begin{subfigure}{\columnwidth}
\begin{lstlisting}[language=Python, basicstyle=\ttfamily\scriptsize, backgroundcolor=\color{gray!5}, escapechar=@, linewidth=\columnwidth, xleftmargin=0pt, xrightmargin=0pt]
    scores_df = pd.read_csv("satscores.csv")
    schools_df = pd.read_csv("schools.csv")
    unique_counties = pd.DataFrame(schools_df["County"].unique(), columns=["County"])
    unique_counties = unique_counties.@\textcolor{red}{sem\_map}@(
        "What is the population of {County} in California? Answer with only the number without commas. Respond with your best guess."
    )
    counties_over_2m = set()
    for _, row in unique_counties.iterrows():
        try:
            if int(re.findall(r"\d+", row._map)[-1]) > 2000000:
                counties_over_2m.add(row.County)
        except:
            pass
    schools_df = schools_df[schools_df["County"].isin(counties_over_2m)]
    merged = pd.merge(scores_df, schools_df, left_on="cds", right_on="CDSCode")
    prediction = int(merged["NumTstTakr"].sum())
\end{lstlisting}
\vspace{-2mm}
\caption{\textnormal{LOTUS program.}}
\label{fig:lotus}
\end{subfigure}

\vspace{0.3cm}

\begin{subfigure}{\columnwidth}
\begin{lstlisting}[language=Python, basicstyle=\ttfamily\small, escapechar=|, mathescape=true, backgroundcolor=\color{gray!5}, numbers = none, xleftmargin=0.5em]
|\(\gamma_{\text{SUM(NumTstTakr)} \rightarrow \text{total\_test\_takers}}\)|
  (|\textcolor{blue!70!black}{satscores}| |\(\bowtie_{\text{cds = CDSCode}}\)|
      (|\(\sigma_{\text{population > 2000000}}\)|
        (|\(\Pi_{\text{\textcolor{red}{ExtractPopulation}(County)} \rightarrow \text{population}}\)|
          (|\textcolor{blue!70!black}{schools}|))))
\end{lstlisting}
\vspace{-2mm}
\caption{\textnormal{HRA representation.}}
\label{fig:hra}
\end{subfigure}

\caption{\textnormal{Example query from the TAG benchmark: ``{\it How many test takers are there at the school/s in a county with population over 2 million?}''. 
(a) LOTUS: expert-written program with explicit execution logic. 
 (b) HRA: declarative algebraic operators.}}
\label{fig:HRA}
\end{figure}

Figure~\ref{fig:HRA} compares how LOTUS \cite{lotus} (using Python semantic analytical programs) and HRA express a query from the TAG benchmark that answers the natural language question: ``{\it How many test takers are there at schools in counties with population over 2 million?}''
This query requires a semantic mapping operation to extract the population of each California county.

In the LOTUS program (Figure~\ref{fig:lotus}), the semantic mapping is performed using a pandas-like API: \texttt{sem\_map} (line 4). Users must specify detailed prompts, handle LLM result parsing, and optimize query execution. For example, the program extracts unique {\tt County} values first to avoid redundant LLM calls (line 3). However, even this expert-written program is suboptimal; performing the join (line 15) before the expensive \texttt{sem\_map} operation would filter counties before the semantic map~\cite{glenn2025blendsql}. 
Such procedural specifications make it difficult for users (both human experts and LLMs) to write optimal analytical programs.

HRA (Figure~\ref{fig:hra}) specifies queries in concise algebraic form. Semantic operations invoke LLMs through user-defined functions (UDFs), represented as function symbols embedded in relational operators (e.g., {\tt ExtractPopulation} in projection). 

\begin{figure}[t]
    \centering
    \includegraphics[width=0.88\linewidth]{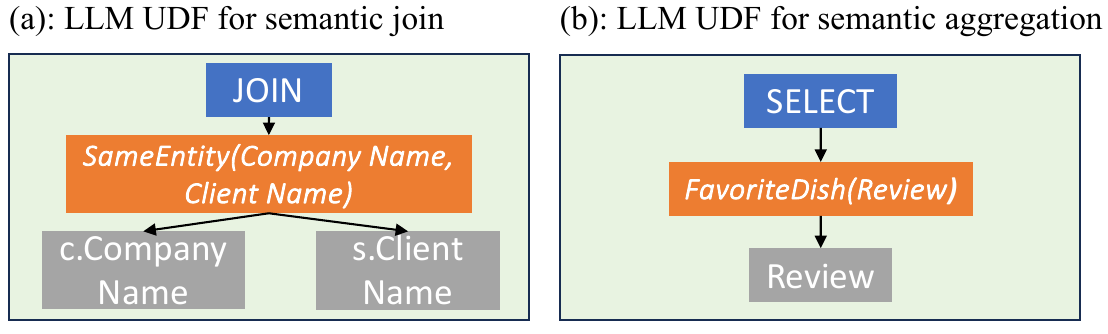}
    \vspace{-2mm}
    \caption{\textnormal{Examples of LLM UDFs in HRA for semantic operations.
    }}
    \label{fig:LLM_model}
    \vspace{-3mm}
\end{figure}

LLM UDFs extend relational operators in multiple ways, as shown in Figure \ref{fig:LLM_model}: they may serve as binary predicates within joins to determine entity equivalence (e.g., {\tt SameEntity} in Figure \ref{fig:LLM_model}a) or as aggregation functions that summarize groups of textual content through semantic processing (e.g., {\tt FavoriteDish} in Figure \ref{fig:LLM_model}b). We formally define an LLM UDF as follows:

\begin{definition} [LLM User-Defined Function] \label{def:udf}
Let $T$ be input relations and $C$ be a subset of columns from $T$, where $T[C]$ denotes the projection of $T$ onto columns $C$.
An LLM-powered UDF $U_M^l$ leverages a language model $M$ to evaluate a natural language expression $l$ that constructs prompts from input $T[C]$. 
The language model $M$ induces a probabilistic distribution $\Pr_M\bigl(y \mid l(T[C])\bigr)$ and returns output $y$ from the output space $\mathcal{Y}$. 
\end{definition}

For example, in Figure~\ref{fig:LLM_model}a, the LLM UDF $U_M^{\textit{SameEntity}}$ evaluates whether two company names refer to the same entity. Given input columns \texttt{Company\_Name} from $T_c$ and \texttt{Client\_Name} from $T_s$, the language model $M$ produces a boolean output ($\mathcal{Y} = \{\texttt{True}, \texttt{False}\}$) for each tuple pair. Consider the input pair  (``\textit{International Business Machines}'', ``\textit{IBM}''); the UDF evaluates it to \texttt{True}. The semantic join then returns all matching pairs: $\{(t_i \in T_c, t_j \in T_s) \mid U_M^{\textit{SameEntity}}(t_i[\texttt{Company\_Name}],$\\$t_j[\texttt{Client\_Name}]) = \texttt{True}\}$.

\begin{table*}[t]
    \caption{\textnormal{
    Summary of HRA operators with their definitions, LLM UDF extensions, and UDF execution algorithms.}}
    \label{tab:example}
    \vspace{-2mm}
    \footnotesize
    \centering
    \begin{tabular}{llll}
  \toprule
    \textbf{Operator} & \textbf{Definition} & \textbf{LLM UDF Extension} & \textbf{UDF Execution Algorithm} \\
    \midrule
    Selection  & $\sigma_\varphi(R)$ & UD-selection predicate ($U_M^l: R \rightarrow bool$)& \makecell[l]{Evaluate $U_M^l$ on distinct inputs, filter and join back with $R$.} \\
    \hline
    Projection & \makecell[l]{$\Pi_{col_1,\dots}(R)$ \\ $\Pi_{f: (col_1, \ldots) \rightarrow col_\text{new}}(R)$ }  & UD-transform function ($U_M^l: R \rightarrow R'$) &   \makecell[l]{Apply $U_M^l$ on distinct inputs to generate value mappings, join with $R$ \\ to produce $col_{\text{new}}$.} \\
    \hline
    Join &  $L \bowtie_{\varphi} R$ & UD-join predicate ($U_M^l: L \times R \rightarrow bool$)&  \makecell[l]{
    Partition join keys into adaptive batches, evaluate batch pairs to find matches \\ for joining $L$ and $R$.} \\
    \hline
    TopK & 
    \makecell[l]{
        $\omega_{\triangleleft(\textit{col}_1, \dots)}(R)$\\
        $\omega_{\triangleleft_k(\textit{col}_1, \dots)}(R)$
    } & 
    UD-compare function ($U_M^l : R \times R \rightarrow bool$)   & \makecell[l]{Compare distinct row pairs via $U_M^l$, aggregate comparison results to obtain ranks, \\join with $R$ to produce top-$k$ results.    
    } \\
    \hline
    Aggregation & \makecell[l]{$\gamma_{{\sf agg}(col)}(R)$ \\ $\gamma_{\mathcal{G}; {\sf agg}(col)}(R)$}&  \makecell[l]{UD-aggregation function ($U_M^l: R \rightarrow R_A$)}&\makecell[l]{Partition by $\mathcal{G}$ (if present), apply $U_M^l$ to each partition to produce results.} \\
  \bottomrule
    \end{tabular}
\end{table*}

Table~\ref{tab:example} lists the operators supported by \omni in HRA. We now demonstrate their semantic extensions using LLM UDFs.

\underline{Selection.} 
The semantic selection incorporates an LLM UDF $U_M^l$ as a selection predicate, which evaluates a natural language condition $l$ using model $M$. This UDF is applied to each row $t_i$, returning a Boolean value indicating whether $t_i$ satisfies the condition, i.e., 
$U_M^l(t_i[C]) \rightarrow \{\texttt{True}, \texttt{False}\}, \forall t_i \in R.$ The operator then leverages the output of $U_M^l$ within the selection predicate to filter table $R$.

\underline{Projection.} 
The semantic projection incorporates an LLM UDF $U_M^l$ as an AI-powered transformation function that extracts new columns through natural language-guided inference using model $M$. This UDF is applied to each row $t_i \in R$, returning a tuple of derived values $\mathbf{y} = \langle y_1, ..., y_k \rangle$ for new columns: $U_M^l(t_i[C]) \rightarrow \mathbf{y}, \forall t_i \in R$.

\underline{Join.} 
The semantic join incorporates an LLM UDF $U_M^l$ as a join predicate, which evaluates pairs of rows from the input relations and predicts a Boolean outcome using model $M$ as: $U_M^l(t_i[C_L], t_j[C_R]) \rightarrow \{\texttt{True}, \texttt{False}\}, \forall t_i \in L, t_j \in R$. The operator returns row pairs for which $U_M^l$ evaluates to \texttt{True}. 

\underline{TopK.} 
The semantic TopK incorporates an LLM UDF as a pairwise comparator $U_M^l(t_i[C],t_j[C]) \rightarrow \{\texttt{True}, \texttt{False}\}, \forall t_i, t_j \in R$, where $U_M^l$ returns a Boolean value indicating whether $t_i$ should be ranked before $t_j$, producing the ordered output table.

\underline{Aggregation.} 
The semantic aggregation incorporates an LLM UDF $U_M^l$ as the aggregation function, enabling semantic summarization over groups of rows. We define such an aggregation function as $U_M^l(t_1[C], \dots, t_k[C]) \rightarrow y_{\text{agg}}, \text{where} \{t_1, \dots, t_k\} \subseteq R$, returning an aggregate result that summarizes the input rows based on the natural language instruction. 

\section{Query Generation} \label{sec:query_generation}
Automatically synthesizing HRA queries from natural language poses three core technical challenges: (1) \textit{semantic-aware schema encoding}—representing database schemas to enable accurate operator selection and target data identification, (2) \textit{compositional query decomposition}—mapping natural language questions to reasoning steps that align with \omni's operator algebra, and (3) \textit{precise UDF synthesis}—generating LLM UDFs that correctly capture query semantics while ensuring executability within HRA. 
Rather than requiring fine-tuning, \omni leverages LLMs' generalization capabilities through an in-context learning framework with three key components (Figure \ref{fig:query-generation-template}):

(1) {\it Semantic Data Model $S$}: We transform the database schema into a semantic representation that encodes column semantics, table relationships, and domain constraints to ensure accurate data usage in the generated query.

(2) {\it Question Decomposition $\mathcal{I}$}: We decompose the user question into reasoning steps. Each step identifies which operator to use and what parameters to apply, continuing until the query task is complete.

(3) {\it Instructions and Exemplars $\mathcal{E}$}: We provide explicit instructions for accurate LLM UDF generation, along with examples that demonstrate how to correctly integrate UDFs within HRA queries.

\begin{figure}[t]
\begin{lstlisting}[basicstyle=\ttfamily\footnotesize, backgroundcolor=\color{gray!5}, frame=single, numbers=none, escapechar=@]
@\textcolor{violet}{\textbf{TASK:}}@ Given the following database schema and the user question, generate a query in HRA grammar.

@\textcolor{teal}{\textbf{[DATABASE SCHEMA]}}@ <semantic_data_model>

@\textcolor{teal}{\textbf{[QUESTION]}}@ <user_question>, <question_decomposition>

@\textcolor{teal}{\textbf{[INSTRUCTIONS]}}@
- Generate queries using HRA grammar syntax
- When using LLM UDFs, follow <llm_udf_instructions>

@\textcolor{teal}{\textbf{[EXAMPLES]}}@ ...
\end{lstlisting}
\vspace{-2mm}
\caption{\textnormal{Prompt template for HRA query generation.}}
\label{fig:query-generation-template}
\end{figure}

We formalize the query generation problem as follows:

\begin{problem}[Query Generation]
\label{prob:query-generation}
Given a natural language question $\mathtt{q}$ and a relational database $D$, \omni constructs a prompt comprising: a semantic data model $\mathcal{S}$, a question decomposition $\mathcal{I}$, and an example set $\mathcal{E}$. We then synthesize an HRA query via in-context learning using an LLM $M_q$:
$$\hat{\phi} = \arg\max_{{\phi}} \mathbb{P}_{M_q}\left({\phi} \mid \mathtt{q}, \mathcal{S}, \mathcal{I}, \mathcal{E}\right)$$
where $\hat{\phi}$ satisfies the following properties:

\begin{enumerate}
    \item \textbf{Syntactic Validity:} $\hat{\phi}$ must conform to the HRA grammar and pass parser validation:
    $$\hat{\phi} \in \mathcal{L}(\mathcal{G}_{\text{HRA}}) \land \textsc{Parse}(\hat{\phi}, \mathcal{S}) \neq \bot$$
    where $\textsc{Parse}$ validates both syntactic correctness and schema consistency with $S$.
    
    \item \textbf{Semantic Correctness:} When executed on database $D$, $\hat{\phi}$ produces results equivalent to the ground truth query $\phi^*$:
    $$\hat{\phi}(D) = \phi^*(D)$$
    When $\phi^*$ is unavailable, semantic correctness is verified through manual inspection or predefined test cases.
    
    \item \textbf{Executability:} $\hat{\phi}$ must execute successfully without runtime errors on the target database:
    $$\textsc{Execute}(\hat{\phi}, D) \neq \bot$$
\end{enumerate}
\end{problem}

We next present the methodology for constructing each component to enable effective query generation.

\subsection{Semantic Data Model} 
To enable effective natural language querying, the LLM must understand the structure and semantics of the underlying database through an appropriate schema representation.
Data Definition Language (DDL) schemas are the most commonly used representation, defining tables and columns through their structural properties.
However, DDL lacks descriptive information such as table and column explanations. To address this limitation, many text-to-SQL systems propose enhanced schema representations \cite{DBLP:conf/coling/WangR0LBCYZYSL25, DBLP:journals/corr/abs-2411-08599}, while benchmarks such as BIRD \cite{DBLP:conf/nips/LiHQYLLWQGHZ0LC23} incorporate database description files to explain abbreviated names and terminologies.

In \omni, we propose a hierarchical \textit{semantic data model} that organizes database information as a YAML configuration, providing a compact and structured representation for both users and LLMs. This model is automatically constructed from database metadata and description files, incorporating three key components:

\textbf{(a) Table Schemas:} Each table and view is represented with its name, description, and column specifications. For each column, we specify:

\begin{itemize}
    \item \textit{Name}: The column identifier from the database metadata.
    \item \textit{Data type}:  The SQL type (e.g., \texttt{INTEGER}, \texttt{VARCHAR}), enabling type-appropriate operations and preventing errors.
    \item \textit{Sample values}: Three representative values from the actual data (e.g., \texttt{2014-09-14 00:00:00.0} for timestamps) that demonstrate formatting conventions and data patterns.
    \item \textit{Description}: A natural language explanation of the column's meaning and constraints (e.g., primary keys). 
\end{itemize}

\textbf{(b) Relationships:} Explicit specifications of table relationships through foreign key constraints and join conditions. These serve two purposes: (1) guiding the LLM toward valid join paths during query generation, and (2) enabling the system to leverage efficient structural joins over expensive semantic joins when appropriate relational constraints exist.

\textbf{(c) Domain Notes:} Domain-specific knowledge capturing business rules and context:

\begin{itemize}
    \item \textit{Computation rules}: Application-specific formulas and calculation methods (e.g., derived metrics).
    \item \textit{Semantic mappings}: Correspondences between natural language expressions and database schema elements (e.g., "revenue" maps to \texttt{price * quantity}).
    \item \textit{Terminology}: Definitions of technical terms and business concepts relevant to the domain.
    \vspace{-0.5mm}
\end{itemize}
This component enables the LLM to apply domain-specific knowledge appropriately, distinguishing specialized business rules from general reasoning. Users can extend the model with additional domain knowledge as needed.

\textbf{Semantic Data Model Filtering.} 
During query generation, \omni filters the semantic data model to reduce context size and improve accuracy. Given a semantic data model $\mathcal{S}$ comprising table schemas, relationships, and domain notes, \omni identifies the relevant subset $\mathcal{S}'$ needed to answer the user's question. 

We employ an LLM to identify relevant columns. The LLM interprets column semantics using their descriptions and sample values, resolves terminology between questions and schema using domain notes, and identifies columns required by computation rules. The filtered model $\mathcal{S}'$ retains all selected columns along with any domain notes referenced during selection.
Beyond explicitly mentioned columns, the LLM also includes entity-identifying columns for semantic reasoning---for example, when the question asks about ``\textit{players from Germany}'' without a nationality column present in the database, it retains player surname and given name to enable LLM-based inference. 
Once relevant columns are identified, we extract the minimal set of relationships that connect the tables containing these columns, ensuring valid join paths in the generated query.

\vspace{-1mm}
\subsection{Question Decomposition}
To generate executable HRA queries from complex user questions, we decompose questions into structured reasoning steps aligned with \omni's operator algebra. Inspired by the \textsc{ReAct}~\cite{react} reasoning paradigm, we provide the LLM with \omni's operator documentation and the semantic data model, then prompt it to iteratively select operators and determine their parameters until the query intent is satisfied.

We format operator documentation as structured descriptions. Each operator specification includes its semantic description, parameter constraints, and usage examples. For instance, \texttt{semantic\_}\\\texttt{projection} describes its capability to extract implicit attributes via LLM inference, specifies parameters including a source relation, input columns, and a target column name, and provides examples illustrating when semantic projection is necessary versus when standard projection suffices.

The decomposition process begins with query intent identification. The LLM analyzes the natural language question to determine its computational goal, such as aggregation (e.g., counting, averaging), filtering (e.g., subset selection), ranking (e.g., top-$k$ retrieval), or existence checking. This identified intent acts as a termination criterion—the LLM continues composing operators until the resulting decomposition produces outputs matching this intent.

At each reasoning step, the LLM selects an operator from the available catalog and determines its parameters by consulting the semantic data model. 
Each step outputs a natural language description of the operator and its application to the data.
For semantic operators, the LLM also provides explicit justification, as these operations are substantially more expensive due to LLM inference costs. This encourages judicious use of semantic extensions—invoking them only when queries need information absent from the schema or require operations beyond standard relational algebra.

Consider the example query in Figure~\ref{fig:HRA}: \textit{"How many test takers are there at the school/s in a county with population over 2 million?"} The question decomposition yields:

\begin{enumerate}
    \item \textit{Semantic Projection:} Extract the {\tt population} for each {\tt county} in Table {\tt schools}.\\
    \textit{Justification:} Population is not stored in the database and must be inferred from county names.
    \item \textit{Select:} Filter Table [\#1] where {\tt population} $> 2{,}000{,}000$.
    \item \textit{Join:} Join Table [\#2] with Table {\tt satscores} on {\tt cds} = {\tt CDSCode}.
    \item \textit{Aggregate:} Sum {\tt NumTstTakr} from Table [\#3] to compute the total number of test takers.
    \vspace{-2mm}
\end{enumerate}

\subsection{Instructions and Exemplars}
Beyond the semantic data model and question decomposition, \omni requires carefully designed in-context instructions and exemplars to ensure accurate UDF synthesis and integration. We employ two complementary strategies:

\textit{(1) Contrastive Prompting.} The natural language expression $l$ in LLM UDFs must precisely capture query semantics. Since \omni automatically constructs prompts from $l$ and input data $T[C]$ (Definition~\ref{def:udf}), ambiguous expressions can produce erroneous or unstructured outputs that fail to execute correctly. To address this, we provide paired examples contrasting precise and ambiguous expressions. For instance, \texttt{is\_R1\_research\_university} (precise) versus \texttt{is\_prestigious} (ambiguous), or \texttt{extract\_population} (precise) versus \texttt{get\_info} (ambiguous). For operators requiring specific output types (e.g., integers for semantic projection), we explicitly instruct the LLM to include type annotations in $l$.

\textit{(2) End-to-End Query Exemplars.} We curate five representative question-query pairs that balance coverage and prompt efficiency. These exemplars span varying complexity (2--6 operators, up to 4 tables) and collectively cover all operator types in \omni's algebra. We include examples for complex query patterns such as comparative aggregations (\texttt{AVG(T1.val) > AVG(T2.val)}) and queries combining multiple semantic operations to guide correct formulation.

\section{Query Optimization} \label{sec:query_optimization}

Generated queries may be suboptimal in execution efficiency. 
Our cost optimization framework jointly accounts for LLM operation costs and relational database operation costs, leveraging symbolic execution \cite{DBLP:conf/icfem/VeanesGHT09} to ensure plan equivalence across transformations.

The optimization process first parses the HRA query into a logical plan, during which \omni's parser validates syntax and verifies that all referenced tables and columns exist in the database.

\begin{definition}[Query Plan] \label{def:plan}
    A query plan $\mathcal{Q} = (V, E, r)$ is a rooted tree where $V = \{\mathsf{op}_1, \mathsf{op}_2, \dots, \mathsf{op}_k\}$ is the set of operator nodes, $E \subseteq V \times V$  is a set of directed edges representing data flow from child to parent operators, and $r \in V$ is the root operator that produces the final query result. 
    Each node in the query tree corresponds to either a relational operator or a semantic operator that incorporates an LLM UDF to perform language-based functions.
\end{definition}

\begin{figure} [t]
    \centering
\includegraphics[width=1\linewidth]{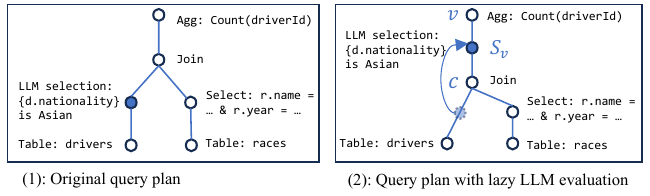}
    \vspace{-6mm}
    \caption{\textnormal{Example: query optimization with lazy LLM evaluation.}}
    \label{fig:lazy}
\end{figure}

\textbf{Problem Statement. }
Given a query plan $\mathcal{Q}=(V,E, r)$ as defined in Definition \ref{def:plan}, its execution cost (estimated runtime) can be represented as:
$$\text{cost}({\mathcal{Q}}) = \sum_{v\in \mathbb{D}}\text{cost}_{\text{sql}}(v) + \sum_{v\in {\mathbb{M}}}\text{cost}_{\text{llm}}(v),$$ 
where $\mathbb{D} \subseteq V$ is the set of relational operators and $\mathbb{M}\subseteq V$ is the set of semantic operators in $\mathcal{Q}$. 

Let $\mathbb{P}(\mathcal{Q})$ denote the set of all query plans that are semantically equivalent to $\mathcal{Q}$. 
The objective of query optimization is to identify an optimal plan that satisfies: 
$$\hat{\mathcal{Q}} = \arg\min_{\mathcal{Q}'\in \mathbb{P}(\mathcal{Q})} \text{cost}(\mathcal{Q'}).$$
Identifying $\hat{\mathcal{Q}}$ is NP-hard, and traditional DBMS cost models and optimization heuristics do not account for the distinct cost characteristics of LLM operations.
For instance, \textit{predicate pushdown}—which moves selection predicates closer to input tables—increases costs for semantic selections by requiring an LLM invocation per input row. Deferring such operators to process smaller intermediate results often reduces LLM invocations and overall cost. 

Our query optimization builds on two insights. First, traditional DBMS optimizations can still reduce  $\text{cost}_{\text{sql}}$ for relational operators in $\mathcal{Q}$. Second, strategically placing LLM UDFs later in the query plan—where they operate on smaller inputs—can reduce $\text{cost}_{\text{llm}}$. 

Based on these insights, \omni introduces a two-phase optimization strategy. Phase one applies standard relational optimizations (e.g., predicate pushdown, join ordering) to minimize intermediate result sizes. Phase two uses a cost-based algorithm to determine optimal LLM UDF placement given the optimized relational plan skeleton. 
Compared with optimization strategies that jointly consider relational and expensive operators \cite{DBLP:journals/corr/abs-2505-14661, DBLP:conf/vldb/ChaudhuriS96}, our approach achieves a more tractable search space by first leveraging mature relational query optimization techniques, then strategically positioning expensive semantic operators.

\textbf{Lazy LLM Evaluation.}
We define the problem as follows:
\begin{problem} [LLM UDF Placement] \label{prob:placement}
Given a query plan $\mathcal{Q} = (V, E, r)$ and a set of semantic operators $\mathbb{M} \subseteq V$ that incorporate LLM UDFs, the task is to find a semantically equivalent plan $\mathcal{Q}'$ that minimizes $\text{cost}(\mathcal{Q}')$ by optimally placing the operators in $\mathbb{M}$ within the query plan.
\end{problem}

Consider an example using a Formula 1 dataset to answer "\textit{How many Asian drivers competed in the 2008 Malaysian Grand Prix?}". The original plan (Figure \ref{fig:lazy} (1)) applies an LLM-based semantic selection on the {\tt nationality} column of the \texttt{drivers} table.
By deferring the execution of this semantic selection (Figure \ref{fig:lazy} (2)), we avoid expensive LLM invocations on all drivers and instead evaluate only the subset of drivers who actually participated in the specified race, thereby reducing overall execution cost.

To solve Problem \ref{prob:placement}, we first define our cost model that unifies LLM invocation costs and relational operation costs.

\begin{definition}[Query Cost Model] \label{def:cost}
For a query plan $\mathcal{Q}$, the execution cost of an operator ${\sf op} \in \mathcal{Q}$ is defined as follows:

Unary operators:
\vspace{-1mm}
\begin{equation}
    \text{cost}({\sf op}_T) = \beta_{{\sf op}} \times |T|
\end{equation}
where $\beta_{{\sf op}}$ is the average cost coefficient for operator ${\sf op}$ to process each tuple, and $|T|$ denotes the cardinality of the input relation $T$.

Binary operators:
\vspace{-1mm}
\begin{equation}
    \text{cost}({\sf op}_{L, R}) = \beta_{{\sf op}} \times f(|L|, |R|)
\end{equation}
where $L$ and $R$ are the left and right input relations, and $f: \mathbb{N} \times \mathbb{N} \rightarrow \mathbb{R}^+$ is a function that depends on the algorithm used by the operator.
\end{definition}

% The function $f(|L|, |R|)$ is algorithm-specific, e.g., $f(|L|, |R|) = |L| \cdot |R|$ for nested loop join and $f(|L|, |R|) = |L| + |R|$ for hash join.
The cost coefficient $\beta_{{\sf op}}$ captures the per-tuple execution cost. For relational operators, this includes disk I/O, CPU processing, and memory access costs~\cite{postgresql_explain}. For semantic operators, $\beta_{{\sf op}}$ estimates model inference overhead based on token consumption (input/output) and per-token processing time. We obtain initial estimates of $\beta_{{\sf op}}$ and operator selectivity through workload sampling for each supported UDF type, which users can optionally refine based on observed performance. For cardinality estimation, we adopt the standard predicate independence assumption \cite{DBLP:conf/sigmod/SelingerACLP79}. 

\begin{algorithm}[t]
\small
\DontPrintSemicolon
\SetKwInOut{Input}{input}\SetKwInOut{Output}{output}
\SetKwFunction{DP}{\textsc{DPCompute}}
\SetKwFunction{Cost}{\text{cost}}
\SetKwFunction{CostLLM}{\textsc{CostLLM}}
\LinesNumbered
\Input{Query plan $\mathcal{Q}=(V, E, r)$, set of semantic operators $\mathbb{M}$}
\Output{Optimal $cost^*[r,\mathbb{M}]$ and corresponding query plan $\mathcal{Q}'$}
\BlankLine

\ForEach{node $v \in \mathcal{Q}$ in postorder traversal}{ \tcp{bottom-up order}
    \uIf{$v$ is a leaf (table initialization)}{
      $cost^*[v,\emptyset] \gets \Cost(v)$
    }
  \ForEach{subset $S \subseteq \mathbb{M}_v$}{ \tcp{$\mathbb{M}_v$: semantic ops in subtree of $v$}
\tcp{$S_v$: subset of $S$ placed immediately under $v$}
    \If{$v$ is unary with child $c$}{ 
      $cost^*[v,S] \gets 
        \min_{S_v \subseteq S} 
          \big( cost^*[c,\, S \setminus S_v] 
             + \Cost_{v,S}(S_v) 
             + \Cost_{S}(v) \big)$}
    \ElseIf{$v$ is binary with children $\ell,r$}{
      $cost^*[v,S] \gets 
        \min_{\substack{S_\ell, S_r, S_v \subseteq S \\ \text{disjoint, } S_\ell \cup S_r \cup S_v = S}}
          \big( cost^*[\ell, S_\ell] 
             + cost^*[r, S_r] 
             + \Cost_{v,S}(S_v) 
             + \Cost_{S}(v) \big)$ 
    }
  }
}
\Return $cost^*[r,\mathbb{M}]$ and the corresponding query plan $\mathcal{Q}'$ \;
\caption{Lazy LLM evaluation}
\label{alg:dp}
\end{algorithm}

Algorithm \ref{alg:dp} presents our approach. We use a memoization table $cost^*[v,S]$ that stores the minimum cost of executing the query subplan rooted at node $v$ with semantic operator set $S$. 
The algorithm processes nodes in bottom-up order (line 1).
For leaf nodes, we initialize $cost^*[v,\emptyset]$ to the table access cost (lines 3-4). For each internal node $v$, we consider all possible subsets $S\subseteq \mathbb{M}_v$ of semantic operators (lines 5-6). 

For any subset $S$, let $S_v \subseteq S$ denote semantic operators repositioned to execute immediately under node $v$ (line 7). We only consider valid placements $S_v$ that preserve plan equivalence. Here, $\text{cost}_S(v)$ represents the execution cost of node $v$ with operator set $S$ in its subtree, and $\text{cost}_{v, S}(S_v)$ represents the cost of executing operators $S_v$ under node $v$.
For unary nodes, we find the optimal subset $S_v \subseteq S$ to place under $v$ that minimizes total cost (lines 8-9). 
For binary nodes, we find the optimal partition of the operators in $S$ among the left child ($S_\ell$), right child ($S_r$), and those placed under $v$ ($S_v$) (lines 10-11). The algorithm returns the minimum cost and optimized plan $\mathcal{Q}'$ (line 12).

The worst-case time complexity of the algorithm is $O(k\cdot 2^m)$, where $k$ is the number of operators in the query plan and $m$ is the number of semantic operators. At each node, for each of the $m$ semantic operators, we have two choices: either reposition it immediately under node $v$ ($S_v$) or leave it in the subtree(s) below.

To determine whether the transformation produces an equivalent plan, we employ symbolic execution \cite{DBLP:conf/icfem/VeanesGHT09, DBLP:conf/sigmod/ChuLWCS17, DBLP:conf/sosp/SchlaipferRLS17, DBLP:journals/pacmmod/YanLH23} with SMT solvers such as Z3 \cite{z3solver} to verify equivalence. 
We model LLM UDFs as uninterpreted functions \cite{z3guide_uf}\footnote{While LLMs may be stochastic in practice, we treat LLM UDFs as deterministic for the purpose of verifying algebraic plan equivalence.}, enabling verification without executing the queries.

We illustrate the algorithm using the example in Figure \ref{fig:lazy}. The algorithm processes nodes in bottom-up order, updating the memoization table $cost^*[v,S]$ for each node $v$ and each subset $S$ of semantic operators. As shown in Figure~\ref{fig:lazy}(2), when processing the aggregation node $v$ with $S=\{\sigma_{llm}\}$, the algorithm compares two possible $S_v$: (1) $S_v = \emptyset$, which keeps the semantic operator $\sigma_{llm}$ within the subtree, or (2) $S_v = \{\sigma_{llm}\}$, which repositions $\sigma_{llm}$ to execute immediately under $v$. If the second option yields a equivalent and more efficient plan, the optimization is applied.

\begin{figure} 
    \centering
    \includegraphics[width=0.9\linewidth]{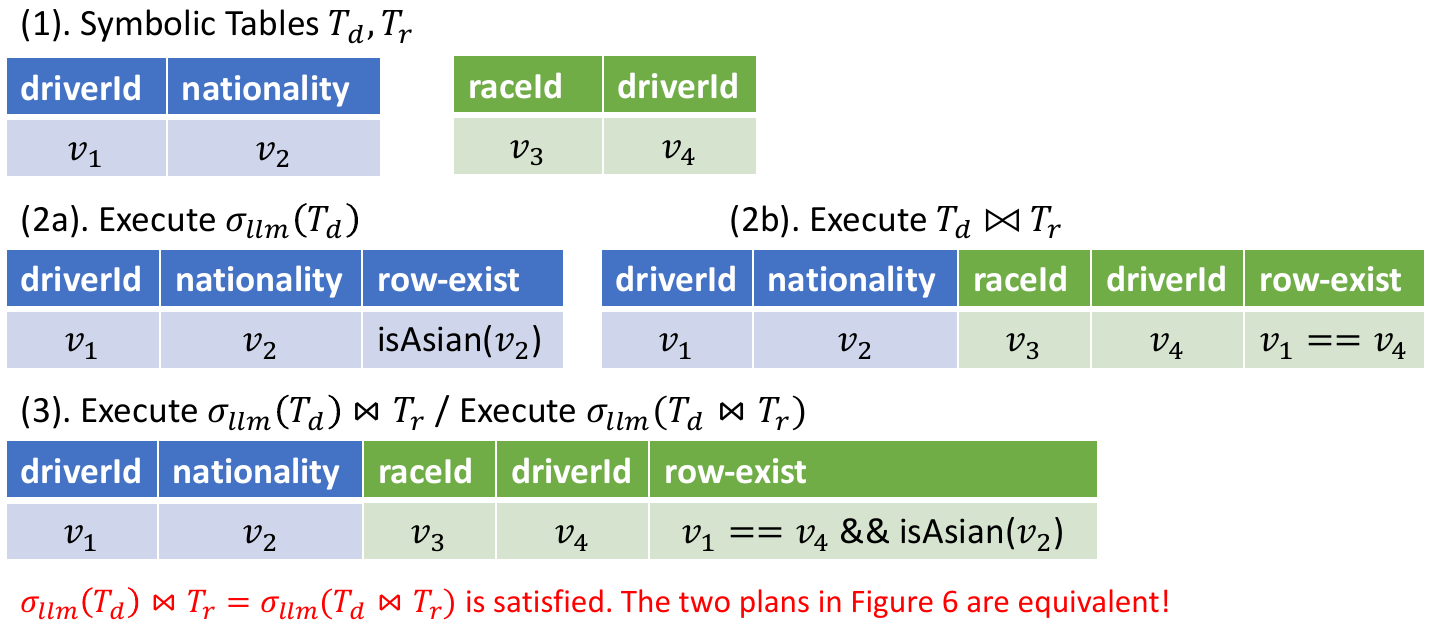}
    \vspace{-2mm}
    \caption{\textnormal{Verification for plan equivalence.}}
    \label{fig:symbolic}
    \vspace{-4mm}
\end{figure}

To verify plan equivalence of the two execution plans, we use symbolic execution to check $\sigma_{llm}(T_d) \bowtie T_r = \sigma_{llm}(T_d \bowtie T_r)$. We first create symbolic tables $T_d$ and $T_r$, each containing a single tuple with cell values represented as symbols $v_1, v_2, v_3, v_4$, and use a \textit{row-exist indicator} for each row to represent filtering conditions (Figure \ref{fig:symbolic}). We then symbolically execute both query plans: (1) The original plan executes the semantic selection on table $T_d$ first, updating the \textit{row-exist} indicator to {\tt isAsian}($v_2$) (Figure \ref{fig:symbolic}(2a)), then executes the join. (2) The optimized plan executes the join first, producing a \textit{row-exist} indicator $v_1 == v_4$ (Figure \ref{fig:symbolic}(2b)), then executes the semantic selection. 
As shown in Figure \ref{fig:symbolic}(3), the two plans yield equivalent outputs.
While this verification is undecidable, the SMT solver provides sound results when it finds equivalence \cite{DBLP:books/aw/AbiteboulHV95}. We revert to the original plan if the solver times out or returns unknown.

\textbf{UDF Rewrite.}
In addition to optimally placing LLM UDFs, we observe that certain LLM UDFs can be eliminated from runtime execution by converting them into equivalent SQL expressions. This applies to UDFs that process each row independently and always produce the same output for the same input, such as certain semantic selections and projections.
For instance, consider an LLM UDF $U_M^l$ with $l:$ \textit{ "Is the user (\{{\tt dob}\}) an Aquarius?"} within a semantic selection. Using the LLM’s knowledge, we could synthesize the equivalent SQL condition
\texttt{strftime('\%m-\%d', dob) BETWEEN '01-20' AND '02-18'}. 

In \omni, we prompt an LLM to analyze the semantics of each LLM UDF $U_M^l$ based on its natural language expression $l$, the input column description, and sampled values. When the LLM determines that a UDF is both stateless and deterministic, it synthesizes an equivalent relational expression that can replace the semantic operator. Otherwise, we retain the original LLM UDF. 
Our evaluation shows that this technique often rewrites UDFs that perform comparisons against real-world constants or rule-based classifications

\textbf{Additional Optimization Opportunities.}
Our framework also creates opportunities for integrating orthogonal physical execution optimizations that could further reduce LLM inference costs.

\textit{Model cascades} \cite{lotus, DBLP:journals/pvldb/KangGBHZ20, DBLP:journals/corr/abs-2310-03094} dynamically route queries to different-sized models based on task complexity, delegating simple tasks to smaller, faster models while reserving larger models for complex reasoning. 
\textit{Proxy models}~\cite{lotus, DBLP:journals/pvldb/YangWHLLW22} pre-filter candidates using cheaper models before invoking expensive models for final predictions, reducing unnecessary expensive inference calls. \textit{Prompt optimization} \cite{liu2025palimpzest} techniques reduce token counts either as preprocessing steps or through dynamic adaptation based on observed performance. 
\textit{Prompt caching} \cite{DBLP:journals/corr/abs-2403-05821, DBLP:conf/mlsys/GimCLSK024, DBLP:journals/corr/abs-2402-05099, DBLP:conf/sosp/KwonLZ0ZY0ZS23} stores and reuses attention states from frequently used prompt segments across similar invocations. This is particularly beneficial when applying the same LLM UDF with shared prompt prefixes to large datasets. We leave the integration of these techniques for future work.

\section{Query Execution} \label{sec:query_execution}
\omni integrates LLM UDFs into database query execution through a UDF executor.  When encountering a semantic operator with LLM UDF $U^l_M(T[C])$, the UDF executor first constructs prompts by instantiating operator-specific templates with the natural language expression $l$ and inputs from $T[C]$. For example, for the semantic filter with LLM UDF $U_M^{\texttt{isAsian}}(\texttt{nationality})$, the executor generates prompts such as: {\it "Does \{Japanese\} satisfy \{isAsian\}? Answer: yes/no" } for each distinct value in the \texttt{nationality} column. The executor then issues parallel LLM calls for these prompts, which return predictions as (\texttt{nationality}, \texttt{prediction}) pairs. 
These pairs are then joined back to the original relation on the \texttt{nationality} column, and the filtered relation flows to the next operator in the query plan.

To minimize redundant LLM invocations, the executor extracts distinct values from input columns before LLM processing for operators where deduplication preserves semantics (selection, projection, join, and top-k).
Table \ref{tab:example} summarizes the UDF execution algorithms for each semantic operator in \omni.

\textbf{Smart-Batching for Semantic Join.}
Existing semantic operator systems \cite{lotus, liu2025palimpzest} often implement semantic joins using a nested-loop approach (Algorithm \ref{alg:nest_join}), which enumerates all possible pairs of distinct join keys from relations $L$ and $R$ (lines 2-3) and evaluates each pair independently using the predicate $U_M^l$ (line 4). This incurs $\mathcal{O}(|K_1| \cdot |K_2|)$ LLM calls, becoming prohibitively expensive for large join relations.

\begin{algorithm}[t]
\small
	\DontPrintSemicolon
	\SetKwInOut{Input}{input}\SetKwInOut{Output}{output}
	\LinesNumbered
	\Input{Relations $L, R$, distinct join keys $K_1, K_2$, LLM UDF $U_M^l$}
	\Output{Set of joinable key pairs $J$} 
    $J \leftarrow \emptyset$ \\
    \ForEach{$k_1 \in K_1$}{
    \ForEach{$k_2 \in K_2$}{
    \If{$U_M^l(k_1, k_2)$ returns \textsc{True}}{
    $J \leftarrow J \cup \{\langle k_1, k_2 \rangle\}$
    }
    }}
	\Return $J$
	\caption{Nested-loop semantic join}
 \label{alg:nest_join}
\end{algorithm}

\begin{algorithm}[t]
\small
	\DontPrintSemicolon
	\SetKwInOut{Input}{input}\SetKwInOut{Output}{output}
	\LinesNumbered
	\Input{Relations $L, R$, distinct join keys $K_1, K_2$, LLM UDF $U_M^l$}
	\Output{Set of joinable key pairs $J$} 
    $J \leftarrow \emptyset$ \\
    $s \leftarrow \text{sample}(K_1, 3) \cup \text{sample}(K_2, 3)$ \\
    $b_1, b_2 \leftarrow M_b(s, |K_1|, |K_2|, U_M^l)$ \tcp{LLM determines batch size}
    $B_1 \leftarrow \text{partition}(K_1, b_1)$ \\
    $B_2 \leftarrow \text{partition}(K_2, b_2)$ \\
    \ForEach{$B_1^i \in B_1$}{
        \ForEach{$B_2^j \in B_2$}{
            $J_{ij} \leftarrow U_M^{l_{\text{batch}}}(B_1^i, B_2^j)$ \tcp{Single LLM call for each batch pair}
            $J \leftarrow J \cup J_{ij}$ \\
        }
    }
	\Return $J$
	\caption{Smart-batching semantic join}
 \label{alg:smart_batch_join}
\end{algorithm}

To improve join efficiency, a naive batching approach combines all $|K_1|$ keys from $L$ and all $|K_2|$ keys from $R$ into a single prompt, asking the LLM to identify and return all matching pairs in one call \cite{DBLP:conf/acl/GlennDWR24}. While this works for simple tasks with short text and small relations, it fails when: (1) the combined context exceeds the LLM's effective context window, (2) complex semantic reasoning degrades with too many pairs in one prompt, or (3) parsing structured output becomes unreliable for large result sets.

\omni\ introduces a smart-batching algorithm (Algorithm \ref{alg:smart_batch_join}) that adaptively determines optimal batch sizes to balance cost and accuracy. The algorithm samples three rows from each relation (line 2) and invokes an LLM $M_b$ that analyzes: (1) \textit{join complexity}—whether the semantic matching through $U_M^l$ requires simple text similarity (e.g., nationality-to-country) or deep semantic reasoning (e.g., matching research methodologies to grant requirements), (2) \textit{context length}—the text length per row based on samples $s$ from each relation, and (3) \textit{data size}—the cardinalities $|K_1|$ and $|K_2|$ of the input relations. Based on these factors, $M_b$ returns batch sizes $b_1$ and $b_2$ (line 3): larger batches (e.g., $b_1 = b_2 = 10$) for simple short-text joins to maximize cost savings, and smaller batches (down to $b_1 = b_2 = 1$) for complex or lengthy text to maintain accuracy. The algorithm then partitions each key set into batches of their respective sizes (lines 4-5) and processes each batch pair with a single LLM invocation (lines 6-9), reducing LLM calls to $\mathcal{O}(\lceil|K_1|/b_1\rceil \cdot \lceil|K_2|/b_2\rceil)$.

Consider joining 7 Formula 1 constructor nationalities with 32 circuit countries using semantic matching (e.g., ``British'' $\leftrightarrow$ ``UK''). The nested-loop approach requires $7 \times 32 = 224$ LLM calls. Smart-batching samples the input data and $M_b$ determines $b_1 = 10$ and $b_2 = 10$, recognizing this as a simple similarity-based matching task suitable for batching. The algorithm partitions the relations into $\lceil 7/10 \rceil = 1$ and $\lceil 32/10 \rceil = 4$ batches respectively. Processing $1 \times 4 = 4$ batch pairs plus 1 batch-sizing call yields just 5 total LLM invocations—a $45\times$ reduction over nested-loop join.
Conversely, for complex joins involving lengthy text or nuanced semantic reasoning, $M_b$ adaptively selects smaller batch sizes to preserve accuracy. By balancing between the nested-loop approach (accurate but expensive) and aggressive batching (efficient but error-prone), smart-batching achieves both cost savings and high prediction quality, ensuring efficient execution across diverse workloads.

\section{Evaluation} \label{sec:evaluation}

\begin{table*}[t]
\footnotesize
\caption{\textnormal{Comparison of baseline approaches.}}
\vspace{-3mm}
\begin{centering}
\begin{tabular}{lp{3.5cm}ccc}
\toprule
\textbf{Methods} & \textbf{System Interface} & \textbf{Query Generation} & \textbf{Query Plan Optimization} & \textbf{Semantic Execution} \\
\midrule
Text2SQL & Natural Language Question & \pcorrect\textsuperscript{\textit{a}} & \wrong & \wrong \\
HQDL & LLM tables + SQL & \wrong & \wrong & \pcorrect\textsuperscript{\textit{b}} \\
BlendSQL & BlendSQL Query & \wrong & \pcorrect\textsuperscript{\textit{c}} & \ \correct \{{\tt LLMValidate}, {\tt LLMMap}, {\tt LLMQA}, {\tt LLMJoin}\} \\
LOTUS & Python (LOTUS API) & \wrong & \wrong & \makecell[c]{ \correct \{{\tt sem\_filter}, {\tt sem\_map}, {\tt sem\_topk}, \\{\tt sem\_join}, {\tt sem\_agg}\}} \\
Palimpzest & Python (Palimpzest API) & \wrong & \makecell[c]{\correct Cascade-style optimization } & \makecell[c]{ \correct \{{\tt sem\_filter}, {\tt sem\_map}, {\tt sem\_join}, {\tt sem\_agg}\}} \\
\textbf{\omni} & Natural Language Question & \correct & \makecell[c]{\correct Lazy LLM evaluation, UDF rewrite}& \makecell[c]{ \correct LLM UDFs incorporated in selection, \\ projection, join, top-$k$, and aggregation} \\
\bottomrule

\multicolumn{5}{l}{\footnotesize \textsuperscript{\textit{a}} Generates only traditional SQL queries. \textsuperscript{\textit{b}}Supports only row-level LLM reasoning; cannot handle semantic joins or aggregations. \textsuperscript{\textit{c}}Rule-based plan transformations.}
\end{tabular}
\end{centering}
\label{table:baseline_comparison}
\vspace{-3mm}
\end{table*}

We conduct an experimental study to evaluate \omni's performance and capabilities. 
We seek to answer the following questions:

\begin{asparaitem}
\item  \textit{RQ1: What capabilities does \omni enable that are challenging with traditional relational querying?} 
We extended the table-augmentation-generation (TAG) benchmark \cite{tag}, which features data-analytics questions that require LLM-based semantic reasoning over relational queries. We systematically evaluated \omni's ability to handle these queries and compared it against a set of baselines, including text-to-SQL systems \cite{DBLP:conf/nips/LiHQYLLWQGHZ0LC23} and state-of-the-art LLM-powered data processing systems \cite{tag, DBLP:conf/acl/GlennDWR24, corr:abs-2408-00884, liu2025palimpzest}.

\item \textit{RQ2: Can \omni automatically generate queries that integrate LLM UDFs, and how reliable is this process?} We enhanced the benchmark with expert-calibrated ground-truth queries and evaluated generated queries for syntactic validity, semantic correctness, and executability. We performed an ablation study examining each component of our framework and its impact on accuracy.
\item \textit{RQ3: How effective is \omni's query optimization?} 
We compared the query execution latency and token usage of optimized versus unoptimized queries.
\item  \textit{RQ4: How effective is the smart-batching algorithm for semantic join?}
We compared our smart-batching algorithm against the nested-loop join baseline, measuring LLM call reduction and execution performance.
\item \textit{RQ5: What is the cost efficiency of \omni?} We analyzed token usage across query generation, optimization, and execution, identifying key factors influencing costs.
\end{asparaitem}

\subsection{Experimental Setup}

\subsubsection{Question Corpus}
We evaluated \omni on TAG~\cite{tag}, a state-of-the-art benchmark for assessing database queries that incorporate LLM reasoning. TAG extends the widely used text-to-SQL benchmark BIRD~\cite{DBLP:conf/nips/LiHQYLLWQGHZ0LC23} by reformulating questions to require LLM capabilities for world knowledge and semantic reasoning over textual columns. For example, in the \textit{California Schools}, a modified query adds an additional clause asking for schools that are in the Bay Area.
The benchmark questions span 5 diverse domains from BIRD: \textit{California Schools}, \textit{Debit Card Specializing}, \textit{Formula One}, \textit{Codebase Community}, and \textit{European Football}. These databases contain up to 13 relations, with the largest reaching 597.8 MB. TAG includes 60 queries covering three question types (\textit{match-based}, \textit{comparison}, and \textit{ranking}) with expert-labeled ground truth, plus 20 \textit{aggregation} queries that perform summarization over textual columns.

\textbf{Extending TAG.}
To enable more comprehensive evaluation,  we extend TAG in two ways. First, we add explicit ground-truth queries. TAG provides only question-answer pairs without intermediate reasoning paths; we engaged three experienced database experts to construct analytical ground-truth queries for each question. This enables direct evaluation of automatically generated queries and independent verification of answer correctness.
Second, we expand the benchmark to include reasoning patterns that commonly arise in real-world AI-powered data analytics~\cite{lotus, liu2025palimpzest, DBLP:conf/nips/LiHQYLLWQGHZ0LC23, DBLP:conf/acl/GlennDWR24, corr:abs-2408-00884} but are not covered in TAG: \textit{semantic joins} (semantic matching across columns beyond string equivalence), \textit{semantic categorization} (classification into categories not represented in the schema), \textit{information extraction} (deriving structured attributes from free text), and \textit{computational reasoning} (applying external knowledge to perform computations).  We added 10 queries per pattern (40 total), introducing multi-hop reasoning, subqueries, and multiple LLM invocations to increase complexity beyond TAG's original scope. We refer to this extended benchmark as TAG+.

\subsubsection{Baselines}
We compared the following data querying systems. 
Table~\ref{table:baseline_comparison} summarizes their characteristics.

\noindent
\textbf{Text2SQL} \cite{DBLP:conf/nips/LiHQYLLWQGHZ0LC23} uses an LLM to generate SQL queries from natural language questions, which are then executed in database engines. We adopt the BIRD prompt format with few-shot examples to generate executable queries with relational operations.

\noindent
\textbf{HQDL} \cite{corr:abs-2408-00884} augments relational databases using LLMs for data imputation, allowing answers to beyond-database questions by filling in missing attributes with LLM-generated values. We specified target attributes and applied the same prompt as \cite{corr:abs-2408-00884} to generate missing values via row-level LLM invocations. With the LLM-augmented tables, we constructed SQL queries to answer the questions.

\noindent
\textbf{BlendSQL} \cite{DBLP:conf/acl/GlennDWR24} extends SQLite to support LLM functions, including {\tt LLMMap}, {\tt LLMQA}, and {\tt LLMJoin}, which enable row-level data transformation, aggregation, and table joins. BlendSQL provides a query interface for users to write and execute queries in BlendSQL syntax. We evaluated it using expert-crafted scripts for each question. BlendSQL applies rule-based optimizations such as predicate pushdown and deferred LLM execution, but does not consider execution costs and uses test sets to verify query equivalence.

\noindent
\textbf{LOTUS} \cite{lotus} implements semantic operators based on the DataFrame abstraction. Each operator is executed using optimized algorithms with statistical accuracy guarantees relative to a "gold algorithm". Users write Python programs using the LOTUS API to manually construct query plans. We implemented Python scripts for each question using LOTUS operators. LOTUS's operator-level physical optimizations trade accuracy for efficiency; we disable them to focus on execution accuracy. LOTUS does not support logical query plan optimization.

\noindent
\textbf{Palimpzest} \cite{liu2025palimpzest} implements semantic operators based on the DataFrame abstraction with a cascades-style optimizer that explores logical and physical implementations to discover a Pareto frontier for quality, cost, and latency trade-offs. Palimpzest requires 
users to manually write programs using its operators. We implemented Python scripts for each question using Palimpzest operators and configured the optimizer to maximize quality ({\tt max\_quality} policy) without cost or latency constraints, ensuring fair comparison of query execution accuracy.

\noindent
\textbf{\omni} automatically generates and executes HRA queries through cost-based optimization with lazy LLM evaluation and UDF rewriting, and employs specialized algorithms for executing semantic operators, such as smart-batching for semantic joins.

\subsubsection{Implementation Details.}
All baseline methods used SQLite3 as the database engine, except LOTUS and Palimpzest, which used Pandas DataFrames. We evaluated all systems using four LLMs: GPT-5~\cite{openai2025gpt5}, Claude Sonnet 4.5~\cite{anthropic2025sonnet45}, Gemini 3~\cite{google2025gemini3}, and an open-weight model Qwen3-256B~\cite{qwen2025technical}. All LLM-based approaches had a 1-hour timeout per query and allowed up to 3 retries to handle API failures and malformed outputs. 
For reproducibility, we set the temperature to 0 unless otherwise stated. 
We set the degree of parallelism for LLM invocations to 10 for all applicable systems (BlendSQL, LOTUS, Palimpzest, and \omni).

\subsection{Baseline Evaluation}

\subsubsection{End-to-End Query Performance}
Table~\ref{table:overall_acc} presents the execution accuracy and token usage of all systems, where we interact with each system via its natural language or code interface as summarized in Table~\ref{table:baseline_comparison}.
For TAG+ questions, we consider results identical to ground truth as correct. For subjective tasks (e.g., summarizing textual comments or ranking schools by perceived promise), we use GPT-5 as an LLM judge to assess whether summarization outputs capture key information effectively, and whether ranking results demonstrate sound reasoning.

LOTUS, Palimpzest, and \omni achieve the best accuracy, as their semantic operator designs better address query requirements. Among these, \omni is the only fully automated approach, while LOTUS and Palimpzest require experts to construct the query pipeline. \omni performs best with Claude Sonnet 4.5, which excels at query generation, and achieves comparable results to LOTUS and Palimpzest with GPT-5 and Gemini 3. With the open-weight LLM Qwen3-256B, \omni reaches 76.7\% accuracy, demonstrating effective performance with open-weight models. 

\omni also achieves comparable token usage to LOTUS and Palimpzest despite requiring additional LLM invocations for automated query generation. This stems from efficient prompt design, query optimization, and smart-batching for semantic joins.

LOTUS shows reduced accuracy on questions requiring LLM parametric knowledge due to its system prompt design, and timeouts on large join inputs from its nested-loop join implementation. Palimpzest shows weaker aggregation performance with Gemini 3, listing high-level information without meaningful summarization.

BlendSQL's lower execution accuracy stems from two factors: (1) batching multiple entries per LLM call increases errors, although this strategy improves token efficiency, and (2) its {\tt LLMMap} function for semantic top-$k$ and aggregation yields lower ranking and summarization performance. 

HQDL achieves relatively low accuracy and high token usage because its row-wise data imputation requires extensive LLM calls to fill missing columns in large relations. This imputation-then-query approach cannot leverage query optimization, causing timeouts on large inputs and failing to support queries requiring cross-row LLM reasoning such as pairwise ranking, aggregation, and joins.

Text2SQL automatically synthesizes SQL queries but cannot handle questions beyond SQL's capabilities. It performs well when reasoning can be expressed through relational operations but poorly on questions requiring LLM inference during execution.

\begin{table}[tp]
\footnotesize
\caption{\textnormal{Execution accuracy and token usage on TAG+. \omni achieves the highest accuracy among automated methods while remaining competitive with manual pipelines ({\footnotesize\textcolor{darkgreen}{$\blacktriangle$}} = best per model).}}
\vspace{-2mm}
\begin{centering}
\begin{tabular}{l|l|lcc}
\toprule
% \rowcolor{mygrey}
\textbf{Methods} & \textbf{Auto.} & \textbf{Models} & \textbf{Exec. Acc. (\%)} & \textbf{Avg. Tokens} \\
\midrule
\multirow{4}{*}{Text2SQL} & \multirow{4}{*}{\correct} & Claude 4.5 Sonnet & 32.5\% & 1781.2 \\
 & &GPT-5 & 34.2\% &  3076.4\\
 & &Gemini 3 & 25.0\% &  4273.8\\
 & &Qwen3-256B & 27.5\% &  1836.3\\
\midrule
\multirow{4}{*}{HQDL} & \multirow{4}{*}{\wrong} & Claude 4.5 Sonnet & 13.3\% & 159126.9 \\
 & &GPT-5 & 12.5\% &  92063.3\\
 & &Gemini 3 & 12.5\% & 268573.7 \\
 & &Qwen3-256B & 11.7\% &  82597.6\\
\midrule
\multirow{4}{*}{BlendSQL} & \multirow{4}{*}{\wrong} & Claude 4.5 Sonnet & 68.3\% & 4214.8 \\
 & &GPT-5 & 65.0\% &  4206.8\\
 & &Gemini 3 & 70.0\% & 3633.3 \\
 & &Qwen3-256B & 65.8\% & 3179.5\\
\midrule
\multirow{4}{*}{LOTUS} & \multirow{4}{*}{\wrong} & Claude 4.5 Sonnet & 80.0\% & 19686.9 \\
 & &GPT-5 & 90.8\%~\textcolor{darkgreen}{$\blacktriangle$} & 31833.6 \\
 & &Gemini 3 & 83.3\%~\textcolor{darkgreen}{$\blacktriangle$} & 46940.2 \\
 & &Qwen3-256B & 82.5\% & 8218.8 \\
\midrule
\multirow{4}{*}{Palimpzest} & \multirow{4}{*}{\wrong} & Claude 4.5 Sonnet & 85.0\% & 39428.6 \\
 & &GPT-5 & 85.0\% & 33939.7 \\
 & &Gemini 3 & 77.5\% & 35385.3 \\
 & &Qwen3-256B & 85.0\%~\textcolor{darkgreen}{$\blacktriangle$} & 41723.5 \\
\midrule
\multirow{4}{*}{\textbf{\omni}} & \multirow{4}{*}{\correct} & Claude 4.5 Sonnet & 89.2\%~\textcolor{darkgreen}{$\blacktriangle$} & 29009.0 \\
 & &GPT-5 & 83.3\% & 35125.1 \\
 & &Gemini 3 & 81.7\% & 45664.4\\
 & &Qwen3-256B & 76.7\% & 25394.8\\
\bottomrule
\end{tabular}
\end{centering}
\label{table:overall_acc}
\vspace{-3mm}
\end{table}

\begin{table}[t]
\centering
\footnotesize
\setlength{\tabcolsep}{4pt}
\renewcommand{\arraystretch}{1.0}
\caption{\textnormal{Error type breakdown on \textsc{Sema-SQL} by model capability.}}
\label{tab:error_analysis}
\vspace{-2mm}
\begin{tabular}{@{}l@{\hspace{8pt}}cccc@{}}
\toprule
\textbf{Error Type} & \textbf{Claude 4.5} & \textbf{GPT-5} & \textbf{Gemini 3} & \textbf{Qwen3-256B} \\
\midrule
\multicolumn{5}{@{}l@{}}{\textit{Instruction Following}} \\
\hspace{1em} Syntax Error & -- & 3.3\% & 1.7\% & 1.7\% \\
\hspace{1em} Parsing Error & 1.7\% & 2.5\% & 1.7\% & 2.5\% \\
\midrule
\multicolumn{5}{@{}l@{}}{\textit{Query Understanding}} \\
\hspace{1em} Misaligned & 1.7\% & 0.8\% & 4.2\% & 5.0\% \\
\hspace{1em} UDF Error & 3.3\% & 2.5\% & 4.2\% & 4.2\% \\
\hspace{1em} Relational Error & -- & 1.7\% & -- & 0.8\% \\
\midrule
\multicolumn{5}{@{}l@{}}{\textit{Parametric Knowledge}} \\
\hspace{1em} Factual Error & 4.1\% & 5.8\% & 6.6\% & 9.1\% \\
\midrule
\textbf{Total Errors} & \textbf{10.8\%} & \textbf{16.7\%} & \textbf{18.3\%} & \textbf{23.3\%} \\
\bottomrule
\end{tabular}
\vspace{-3mm}
\end{table}

\begin{figure*}[tp]
    \centering
            \begin{minipage}[t]{0.3
    \textwidth}
        \centering
        \includegraphics[width=0.76\linewidth]{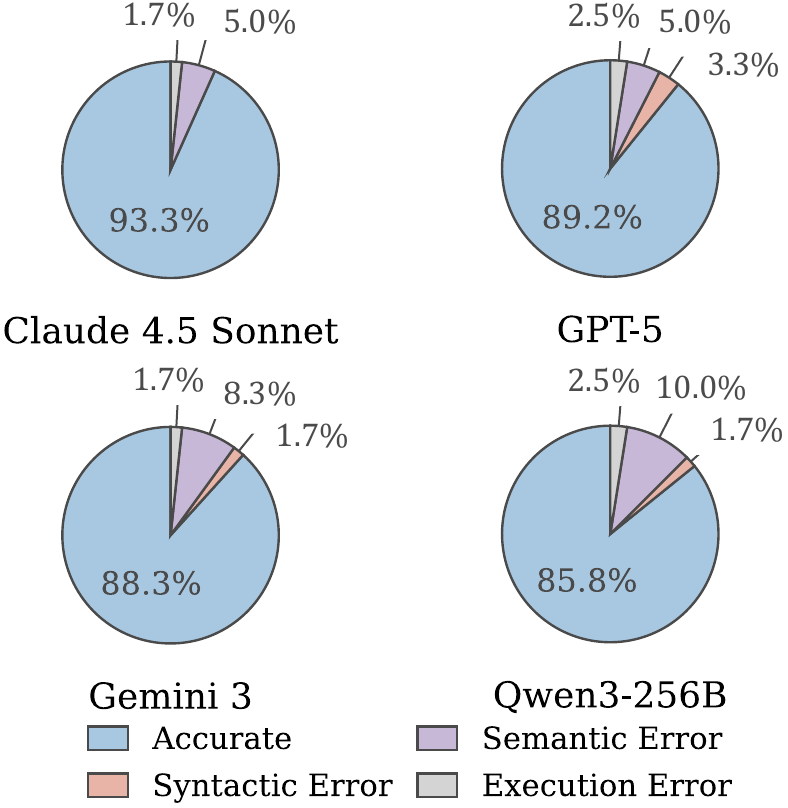}
        \vspace{-3mm}
        \caption{\textnormal{Query generation accuracy.}}
        \label{fig:error}
    \end{minipage}
        \hfill
    \begin{minipage}[t]{0.34
    \textwidth}
        \centering
        \includegraphics[width=\linewidth]{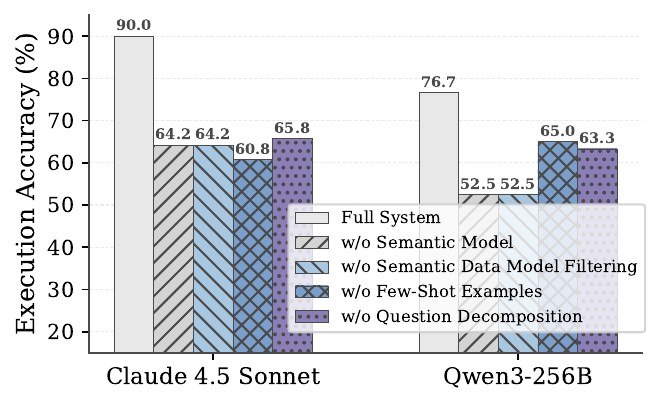}
        \vspace{-9mm}
        \caption{\textnormal{Ablation study of query generation components in execution accuracy.
        }}
        \label{fig:error}
    \end{minipage}
    \hfill
    \begin{minipage}[t]{0.35\textwidth}
        \centering
        \includegraphics[width=\linewidth]{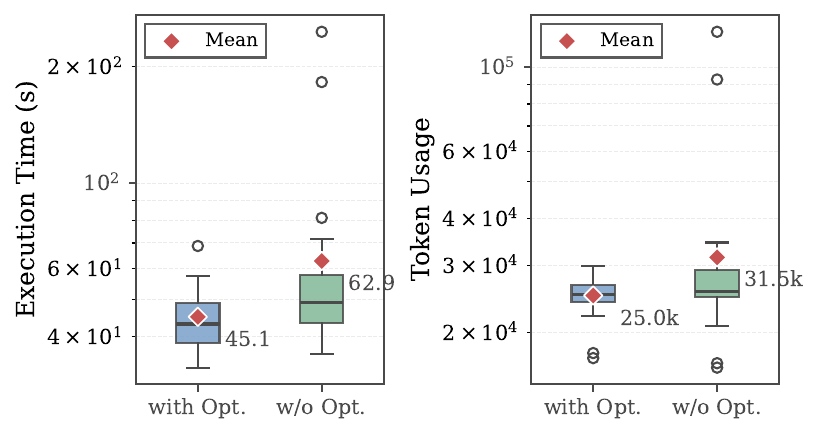}
        \vspace{-9mm}
        \caption{\textnormal{Execution time and token usage comparison with and without query optimization. 
        }}
        \label{fig:token}
    \end{minipage}
\vspace{-2mm}
\end{figure*}

\subsubsection{\omni Error Analysis}
Table \ref{tab:error_analysis} categorizes errors across three core model capabilities—\textit{instruction following}, \textit{query understanding}, and \textit{parametric knowledge}—revealing how \omni's design remains effective across diverse LLMs while highlighting opportunities for enhancement of open-weight models.

Instruction following errors (1.7\%--5.8\%) reflect adherence to grammar and output specifications: \textit{Syntax errors} involve grammar violations such as column hallucination or invalid UDF generation, while \textit{Parsing errors} indicate a failure to produce correctly structured outputs. Claude 4.5 Sonnet excels with only 1.7\% errors, while GPT-5 shows the highest rate at 5.8\%.

Query understanding errors (5.0\%--10.0\%) arise from failures in question comprehension and query synthesis: \textit{Misaligned errors} occur when the generated query does not match the natural language intent (e.g., counting wrong objects or missing filters); \textit{UDF errors} stem from incorrect understanding of required input columns or inability to express operations accurately; \textit{Relational errors} involve flawed predicates or join conditions. Qwen3-256B exhibits the highest rate at 10.0\%, suggesting it may benefit from task-specific fine-tuning to improve query understanding capabilities.

Parametric knowledge errors (4.1\%--9.1\%) reflects world knowledge gaps and the quality of knowledge inference. Qwen3-256B's 9.1\% rate is 2.2$\times$ higher than Claude's 4.1\%, underscoring that retrieval augmentation or continued pretraining on domain-specific corpora could further elevate open-weight models to competitive performance levels.

\subsection{\omni Performance Evaluation}

\subsubsection{Accuracy of Query Generation}

We evaluated query generation accuracy across three dimensions: syntactic validity, semantic correctness, and executability (Problem \ref{prob:query-generation}). For semantic correctness, we compared the execution results of the LLM-generated queries against those of the ground-truth queries.

Figure \ref{fig:error} demonstrates that \omni achieves highly reliable query generation across all models, with accuracy consistently above 85\%. Claude 4.5 Sonnet achieves the highest accuracy at 93.3\%, while Qwen3-256B achieves 85.8\%. 
Syntactic errors (e.g., parsing failures and hallucinated columns) are detected by the \omni parser, whereas execution errors (e.g., value/type mismatches and other database exceptions) are detected by the UDF executor. When such errors are identified, \omni retries query generation using the resulting error message as feedback, up to a maximum retry limit. The remaining errors are predominantly semantic—queries that fail to capture precise intent—typically due to ambiguous question interpretations. Improvements in ambiguity resolution and schema grounding offer promising directions to address these.

\subsubsection{Ablation Study of Prompt Components}

To explore the effect of the key components in query generation—(1) semantic data model, (2) question decomposition, and (3) in-context examples—we conducted an ablation study on Claude 4.5 Sonnet (best performing) and Qwen3-256B (lowest performing) to assess the impact of each component on execution accuracy.

Qwen3-256B is most sensitive to database representation. Removing either the semantic data model representation (compared to BIRD's default schema format \cite{DBLP:conf/nips/LiHQYLLWQGHZ0LC23}) or semantic data model filtering decreases accuracy by 24.2\%. Without these components, generated queries exhibit higher rates of column hallucinations and incorrect operator application. This sensitivity likely stems from Qwen3-256B's shorter context length ($\sim$32K tokens) and lower capability, making compact, relevant schema representation essential for performance.

Claude 4.5 Sonnet benefits most from in-context examples. 
Question decomposition proves crucial for both models. Without it, direct mapping from natural language to HRA queries results in missing computation steps and incorrect semantic operator usage, highlighting the importance of explicit reasoning steps and detailed operator documentation for accurate query generation.

\subsubsection{Effectiveness of Query Optimization} 
We assessed the impact of query optimization in \omni. Of the 120 benchmark queries, optimization transformed 33 query plans. The remaining queries were already optimal from the generation phase or contained LLM UDFs that could not be rewritten.

Figure \ref{fig:token} reports query execution latency (in seconds) and LLM invocation costs for these queries under Claude 4.5 Sonnet. With optimization enabled, we achieved both runtime and cost reduction: average query execution time drops from 62.9 seconds to 45.1 seconds (28\% reduction), while average token consumption decreases from 31.5K to 25.0K (21\% reduction).

Analyzing the optimized plans versus their unoptimized variants reveals how each technique contributes to these improvements. Our two-phase optimization with lazy LLM evaluation optimally places LLM UDFs within query plans so they process smaller intermediate results, reducing LLM invocation costs.  For instance, in the TAG+ query shown in Figure \ref{fig:lazy}, our optimization repositions the UDF after the join predicate, reducing the input from 73 to 42 rows after the join filters irrelevant data. In another TAG+ query containing both a semantic selection and a semantic join, our optimization places the selection before the join, as the selection provides higher selectivity and lower per-row execution cost.

UDF rewriting provides complementary benefits by replacing LLM invocations with equivalent SQL expressions. For example, the TAG+ query contains a semantic selection to determine if a county is in the Bay Area. Without optimization, this would require 57 separate LLM calls (one per row). UDF rewriting recognizes that this semantic condition can be expressed as a SQL predicate: \textit{County IN ('Alameda', 'Contra Costa', 'Marin', 'San Francisco', 'San Mateo', 'Santa Clara', 'Santa Cruz', 'Solano', 'Sonoma')}. This transformation eliminates all 57 LLM calls during query execution, reducing both execution time and costs for this operation.

\subsubsection{Smart-Batching for Semantic Joins} \label{sec:exec_result}

\begin{figure}[t]
    \centering \includegraphics[width=0.98\columnwidth]{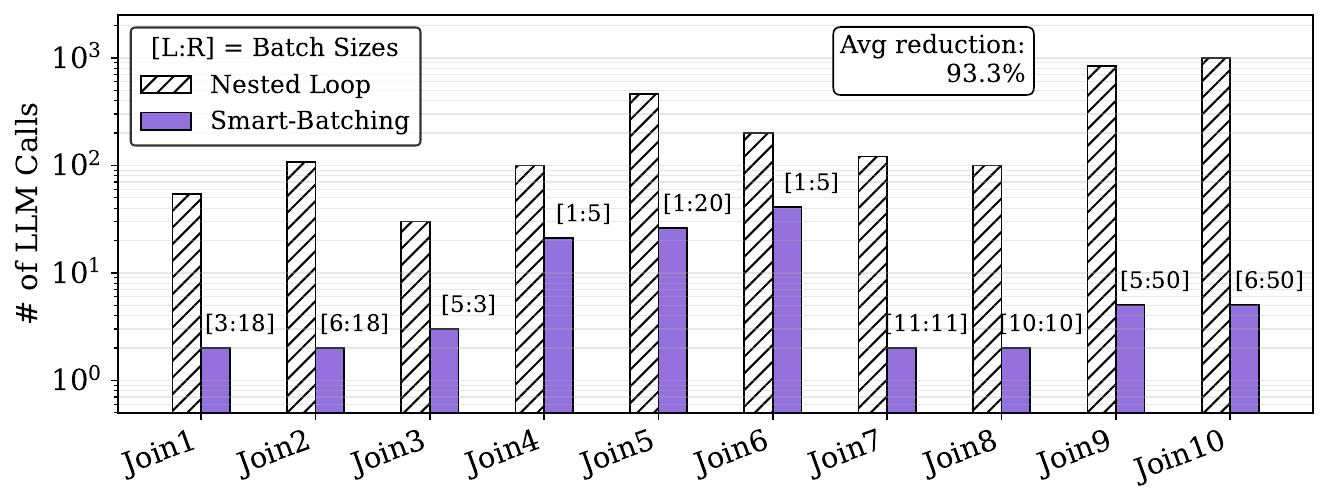}
    \vspace{-4mm}
    \caption{\textnormal{Comparison for semantic join algorithms.}}
    \label{fig:llm_calls}
\end{figure}

We evaluated our smart-batching against the nested-loop baseline widely used in existing systems \cite{lotus, liu2025palimpzest, bigquery}, including LOTUS and Palimpzest. Figure \ref{fig:llm_calls} shows that across 10 benchmark join queries using Claude 4.5 Sonnet, smart-batching reduces LLM calls by 93.3\% while maintaining 100\% execution accuracy (other models show similar results).

The algorithm's adaptive behavior explains its effectiveness across diverse join scenarios. Figure \ref{fig:llm_calls} annotates batch sizes as [L:R], indicating the number of records processed per call from the left and right relations, respectively. For simple similarity-based joins with compact contexts—nationality-to-country matching (Joins 1, 2) and team-to-country mapping (Joins 7, 8)—the smart-batching algorithm selects large batch sizes (often processing entire tables in a single prompt), yielding up to 100$\times$ cost reductions without accuracy loss.
Conversely, for joins requiring deep semantic reasoning—post-to-comment matching (Joins 4, 6) or comment-to-tag mapping (Join 5)—the algorithm dynamically reduces batch sizes or processes records individually. This adaptive fallback prevents the accuracy degradation that naive batching as in BlendSQL \cite{DBLP:conf/acl/GlennDWR24} would cause.
For intermediate-complexity joins (Joins 3, 9, 10), batch sizes also scale proportionally with context lengths and problem complexity.

Join 10 (1,000+ pairwise comparisons) exposes the nested-loop baseline's scalability limitations: it causes end-to-end execution timeouts in LOTUS and incurs high execution overhead in Palimpzest due to excessive latency, while smart-batching consistently completes successfully with significantly lower overhead.

\subsubsection{Cost Analysis}
We reported \omni's token usage across three phases of query processing (query generation, optimization, and execution) and examined the key factors influencing consumption. Table \ref{tab:token_cost_analysis} shows that each query consumes 25K--46K tokens on average, with query generation accounting for 60--75\% of total cost, execution consuming 20--35\%, and optimization using only 3--8\%.

Query generation dominates token usage by processing the complete semantic data model, which scales with schema size. For instance, the {\it European Football} dataset (13 relations) consumes 4.9$\times$ more tokens than {\it Debit Card Specializing} (5 relations).
Query optimization incurs low token consumption (under 2K tokens on average), requiring only UDF characteristic analysis and SQL synthesis from natural language expressions and input samples.
Query execution costs correlate directly with input data volume and semantic processing context—larger inputs result in proportionally higher token consumption.

\begin{table}[tp]
\centering
\caption{\textnormal{Average token cost efficiency per query for \omni.}}
\label{tab:token_cost_analysis}
\vspace{-2mm}
\footnotesize
\begin{tabular}{@{}lrrrr@{}}
\toprule
\textbf{Phase} &  \textbf{Claude Sonnet 4.5} & \textbf{GPT-5} & \textbf{Gemini 3} & \textbf{Qwen3-256B} \\
\midrule
Query Generation & 18331.9 & 20813.4 & 25021.1 & 16774.4\\
Query Optimization  & 1609.0 & 2014.7 & 1347.2 & 1333.1\\
Query Execution  & 9068.1 & 12297.0 & 19296.1 & 7287.3\\
\midrule
\textbf{Total } & 29009.0 & 35125.1 & 45664.4 & 25394.8\\
\bottomrule
\end{tabular}
\vspace{-3mm}
\end{table}
\section{Related Work} \label{sec:related}
\textbf{\textit{Hybrid Question Answering.}}
Hybrid question answering addresses queries over both structured and unstructured data. LLMs excel at semantic processing but lack correctness guarantees for precise operations. SQL provides efficient, verifiable operations but cannot handle semantic ambiguity or reasoning across heterogeneous contexts. 
Hybrid systems combine structured relations with semistructured information including free-form text \cite{DBLP:conf/emnlp/ChenZCXWW20, naacl:LiuXTSYL24, lotus}, LLM knowledge \cite{corr:abs-2408-00884, DBLP:conf/edbt/0002CP24, naacl:LiuXTSYL24}, and knowledge bases \cite{stark}.
Retrieval-augmented generation (RAG) extends to tabular data \cite{DBLP:conf/naacl/HerzigMKE21, DBLP:conf/acl/ChenZR24} and multi-hop reasoning \cite{DBLP:conf/emnlp/ChenZCXWW20}, but remains limited to point-wise retrieval and simple filtering. This paper explores richer queries requiring coordinated relational operations and LLM reasoning \cite{tag}.

\textbf{\textit{Semantic Operator Systems.}} 
Recent work has explored integrating LLM functionalities to extend the scope of relational data processing. Several commercial DBMSs \cite{bigquery, databrick, snowflake, redshift, bigquerydataframe} enable LLM inference within SQL execution.
Academic research has also proposed systems that expose LLMs via declarative operator APIs to extend query languages. BINDER \cite{DBLP:conf/iclr/ChengX0LNHXROZS23} provides a unified API for integrating LLM reasoning into languages like SQL and Python, while BlendSQL \cite{DBLP:conf/acl/GlennDWR24} introduces an SQLite extension with LLM functions for reasoning over multi-table databases. Several systems have implemented optimized LLM operations for specialized data analytical tasks. ZenDB \cite{corr:abs-2405-04674}, EVAPORATE \cite{pvldb:AroraYENHTR23},  Galois \cite{DBLP:conf/edbt/0002CP24}, and HQDL \cite{corr:abs-2408-00884} use LLMs to extract unstructured and semi-structured data into relational formats to support downstream analysis. SUQL \cite{naacl:LiuXTSYL24} augments SQL with LLM operators to summarize free-form text and support conversational question answering. Recent systems including DocETL \cite{DBLP:journals/pvldb/ShankarCSPW25}, Aryn \cite{DBLP:journals/corr/abs-2409-00847}, LOTUS \cite{lotus}, Palimpzest \cite{corr:abs-2405-14696, DBLP:journals/corr/abs-2505-14661, liu2025palimpchat}, and ThalamusDB \cite{DBLP:journals/pacmmod/JoT24} provide declarative semantic operators for AI workloads over unstructured data. They integrate LLMs as native operators in data pipelines, enabling users to perform complex semantic operations—such as extraction, classification, transformation, and joining—on text, documents, and multi-modal data through high-level abstractions.

These systems often require manual query composition, and only a few works \cite{naacl:LiuXTSYL24,DBLP:conf/iclr/ChengX0LNHXROZS23, liu2025palimpchat} attempt automated synthesis. However, these approaches support limited operator types and lack a comprehensive solution for end-to-end query processing.
\omni provides fully automated processing across all stages—query generation, optimization, and execution—for complex queries requiring both relational and semantic processing.

\textit{\textbf{Semantic Query Optimization Approaches.}}
Existing frameworks take different approaches to optimizing the execution of semantic queries:
\textit{(1) Proxy-based execution optimization.} 
LOTUS~\cite{lotus} predefines a set of "gold algorithms"—the ideal but expensive implementations of semantic operators using powerful models. It then optimizes execution by approximating them with cheaper alternatives. 
Similarly, \cite{DBLP:journals/pvldb/YangWHLLW22} injects lightweight proxy models before expensive UDFs to filter unlikely inputs early.
While these approaches effectively optimize physical operator execution, they require users to manually design the query pipeline, where join ordering and operator placement critically impact execution costs.
\textit{(2) Rule-based optimization.} Systems like~\cite{bigquery, DBLP:conf/acl/GlennDWR24, naacl:LiuXTSYL24, DBLP:journals/corr/abs-2508-05002} apply predetermined transformation rules such as predicate pushdown and deferred LLM execution, regardless of data characteristics. This approach has two key limitations: it may miss valuable optimization opportunities, and certain rules may not always yield optimal performance. 
\textit{(3) LLM-based logical plan rewriting.} DocETL \cite{DBLP:journals/pvldb/ShankarCSPW25} uses LLM agents to rewrite query plans, employing validation agents to select plans based on accuracy metrics. Aryn \cite{DBLP:journals/corr/abs-2409-00847} translates natural language queries to semantic plans via Luna, relying on human verification through plan inspection and modification. Nirvana~\cite{zhu2025relationalsemanticawaremultimodalanalytics} applies LLM-based rewriting with natural-language rules for logical optimization and employs a cost-aware physical optimizer for LLM backend selection. However, these approaches require additional verification mechanisms due to the inherent unreliability of LLMs.
\textit{(4) Cost-based optimization.} Palimpzest employs Abacus \cite{DBLP:journals/corr/abs-2505-14661}, a Cascade-style optimizer that estimates operator cost, latency, and quality through sampling. Abacus performs Pareto optimization to balance multiple objectives (e.g., maximizing quality under cost constraints). In contrast, \omni leverages traditional DBMS optimizations and a cost-based algorithm to determine the optimal placement of expensive LLM UDFs, achieving efficient plans through a tractable search space.

\section{Conclusion} \label{sec:conclusion}
We present \omni, a system that automatically translates natural language questions into efficient queries combining relational operations with LLM-based semantic reasoning. \omni introduces Hybrid Relational Algebra (HRA), which extends relational algebra with semantic operators via LLM UDFs. 
The system provides an end-to-end pipeline: in-context learning for query generation, cost optimization accounting for expensive LLM operations, and efficient execution algorithms including smart-batching for semantic joins. Experiments show that \omni significantly enhances querying capabilities compared to baseline approaches.

\bibliographystyle{abbrv}
\bibliography{ref}

\end{document}